\documentclass[twoside,10pt]{llncs}
\usepackage{a4wide}
\usepackage{amsmath, amssymb}
\usepackage[isolatin]{inputenc}
\usepackage[british]{babel}
\usepackage{pifont}
\usepackage{epsfig} 
\usepackage{theorem}

\newtheorem{dfn}{Definition}[section] 
\newtheorem{prop}[dfn]{Proposition}
\newtheorem{theo}[dfn]{Theorem}
\newtheorem{cor}[dfn]{Corollary} 
\newtheorem{ex}[dfn]{Example}
\newtheorem{rem}[dfn]{Remark}
\newtheorem{lem}[dfn]{Lemma}
\newtheorem{fact}[dfn]{Fact}

\newtheorem{prov}[dfn]{Proviso}

\newcommand{\bigmid}{\;\big|\;}

\newcommand{\tw}{\textup{tw}}
\newcommand{\bw}{\textup{bw}}

\newcommand{\mtw}{\textup{mtw}}
\renewcommand{\r}{\textup{r}}
\newcommand{\w}{\textup{w}}

\newcommand{\nw}{\textup{node-w}}
\newcommand{\vfnw}{\textup{VF-node-w}}
\renewcommand{\c}{\textup{c}}

\newcommand{\vftw}{\textup{VF-tw}}
\newcommand{\Part}{\textup{Part}}

\newcommand{\restrict}{\upharpoonright}
\newcommand{\mP}{\mathcal P}
\newcommand{\mS}{\mathcal S}
\newcommand{\mB}{\mathcal B}
\newcommand{\mF}{\mathcal F}
\newcommand{\Pfk}{\textup{Part}_f^k}
\newcommand{\Pmtwk}{\textup{Part}_\mtw^k}
\newcommand{\Ssing}{\mS_{\textit{sing}}}
\newcommand{\Ptwk}{\textup{Part}_\tw^k}
\newcommand{\gf}{\textup{GF}(4)}

\author{Isolde Adler}

\title{Games for width parameters and monotonicity}
\titlerunning{Games for width parameters and monotonicity}
\institute{Department of Informatics, University of Bergen, N-5020 Bergen, Norway\\
\email{isolde.adler@ii.uib.no}}
\begin{document} 
\maketitle


\begin{abstract}
We introduce a search game for two players played on 
a \emph{scenario} consisting of a
ground set together with a collection of feasible partitions.
This general setting allows us to obtain new characterisations of
many width parameters such as rank-width and carving-width of graphs, 
matroid tree-width and GF$(4)$-rank-width. 
We show that the \emph{monotone} game variant corresponds
to a tree decomposition of the ground set along feasible partitions.
Our framework
also captures many other decompositions into `simple' subsets of the ground set,
such as decompositions into planar subgraphs.

Within our general framework, we take a step towards characterising 
monotone search games.
We exhibit a large class of \emph{monotone} scenarios, i.e.~of 
scenarios where the game and its monotone variant
coincide. As a consequence, determining the winner is
in NP for these games.
This result implies monotonicity
for all our search games,
that are equivalent to branch-width of a submodular function.

Finally, we include a proof showing that
the matroid tree-width of a graphic matroid is not larger
than the tree-width of the corresponding
graph. This proof is considerably shorter than the original proof and
it is purely graph theoretic.

\end{abstract}


\section{Introduction}
Search games were introduced by Parsons and Petrov 
in \cite{parsons76,parsons78,petrov82} and since then gained 
a lot of 
interest both in computer science and discrete 
mathematics \cite{AignerF84,BodlaenderT04,biesey91,gromar06,Obdrzalek06,HunterK08,FominGK08}.
In search games on graphs, a fugitive and a 
set of searchers move on a graph, according to 
some rules. The searchers' goal is to capture the
fugitive, and the fugitive tries to avoid capture
indefinitely. Depending on the rules, different
variants of search games arise.

These games have applications in various areas.
On one hand, they are used to model a variety of
real-life problems such as searching a lost person in a
system of caves \cite{parsons76}, clearing tunnels that are contaminated
with gas \cite{LaPaugh93}, and modeling bugs in
distributed environments \cite{FranklinGY00}. On the other hand,
search games are strongly related to graph structure theory, especially to width parameters,
such as tree-width \cite{bienstock91,SeymourT93}, 
path-width \cite{biesey91}, cutwidth \cite{MakedonS89}, 
directed tree-width \cite{jrst01},
and many others. 
They provide a better understanding of the parameters, since a winning strategy
for the cops is a witness for the parameter being small, 
whereas a winning strategy for the robber is a obstruction 
for a small parameter. A game characterisation of a width parameter helps 
in finding examples, and in many cases, games allow for a
polynomial time approximation for the problem of deciding whether the
corresponding parameter is bounded by some fixed integer $k$.

Yet not all width parameters have game characterisations.
Our general framework allows us to fill a large part of this gap.
We introduce 
a game parameter that is within a factor of $3$ of branch-width of a submodular function. 
In particular, we obtain games equivalent to 
rank-width~\cite{oumsey05}, carving-width~\cite{SeymourT94},  
and GF$(4)$-rank-width and bi-rank-width~\cite{kante07}, and we give an exact game characterisation
of matroid tree-width~\cite{HlinenyW06} and for tree-width 
of directed graphs as introduced by
Reed in~\cite{reed99}.  
Moreover, we characterise all our game parameters by a parameter
defined via a `tree decomposition'. We hope that these new
characterisations give a deeper insight, especially into
the newer notions such as rank-width,
GF$(4)$-rank-width and bi-rank-width, and maybe even help
solving Seese's conjecture~\cite{CourcelleO07}.

One very important and desirable property of search game
is monotonicity. Intuitively, a search game is \emph{monotone}, 
if, in the case that the searchers
can catch the fugitive, the searchers can catch him
without having to search a previously searched area again.

If a search game is monotone, this gives us a
polynomial space certificate for proving that determining the 
winner 
in NP, because we can restrict ourselves to
monotone search strategies only. But not all search 
games are monotone, and
although monotonicity is a well-studied property,
until now there is no general method for
distinguishing monotone games from non-monotone.
Actually, some of the most involved techniques
in the area of graph searching were developed for showing
monotonicity~\cite{LaPaugh93,biesey91,seytho93,KirousisP86,FraigniaudN08,MazoitN07}. 
Recent developments in this direction contain new results 
concerning monotonicity of search games on 
directed graphs and hypergraphs, and here many important questions remain 
unsolved~\cite{jrst01,BerwangerDHK06,HunterK07,KreutzerO08,adl04,AdlerGG07}.
See \cite{Fomthi08} 
for a survey. 
Since monotonicity has attracted so much interest, a natural
question arisies:
Can we characterise monotone search games?

In this paper, we consider this question and give results that provide
a step towards its resolution. We introduce
a general framework for a variant of search games where the
fugitive is visible, and the searchers -- in our case there
is just one \emph{captain} -- try to corner him by building barriers.
In our framework, the fugitive is a robber moving on elements of a finite set $A$, 
and the captain has a collection $\mP$ of `feasible' partitions of
$A$. In each round, the captain chooses a new feasible partition,
and rebuilds the barriers accordingly. The robber tries to
escape, but his moves are limited by the (partial) barriers that persist
during the process of rebuilding. We also introduce a collection
$\mS$ of `simple' subsets of $A$. These are subsets
of $A$ where catching the robber is trivial. We say that the
captain wins, if she manages to corner the robber in a simple
subset. 
We call such a pair $(\mP,\mS)$ (satisfying some natural properties) 
a \emph{scenario} on $A$. 
Generalising
ideas of Amini et al.~\cite{ammanith08},
we introduce \emph{weakly submodular} scenarios, and we show that
the games on weakly submodular scenarios are monotone.

We keep the assumptions on the scenario
very weak, in order to shed light on the conditions 
that imply monotonicity. 
Moreover, since we can decompose any scenario, we obtain 
decompositions of graphs into any kind of `simple' subgraphs, 
such as planar graphs or $H$-minor free-graphs. In this way
it should also be possible to find applications in future research.

Our framework also yields a game characterising
tree-width of graphs (our game parameter
is one less that the number of cops necessary to
catch the robber in the robber-and-cops game
\cite{seytho93} characterising tree-width).

Let us give an intuition of our game for tree-width:
When specialising matroid tree-width 
to graphs, the notion yields a simple
equivalent definition \cite{HlinenyW06,HlinenyW08} of graph 
tree-width:
A \emph{tree decomposition} of a graph $G$ is then merely a tree $T$, 
whose leaves are labeled by the edges of $G$. Every internal tree 
node $t\in V(T)$ defines a partition $P_t$ on the edges $E(G)$: the partition 
corresponding to the leaf labels of the connected components of $T\setminus t$. 
The \emph{width} of $t$ is
then the number of vertices of $G$ on the \emph{boundaries} of

the sets in $P_t$ (vertices incident with edges from different partition sets). 
As usual, the width of a tree decomposition is
the maximum of the widths of its tree nodes, and the tree-width
of $G$ is the minimum possible width of a tree decomposition of $G$.
For every integer $k\geq 0$ we let $\mP_k$ denote the collection 
of partitions of $E(G)$ with boundaries of size at most $k$.
In the corresponding game, the robber moves on edges of $G$, 
and the captain chooses partitions from $\mP_k$. The
captain has to catch the robber by cornering him on an edge. 

The paper is organised as follows.
We begin by introducing the captain and robber game 
on scenarios in Sect.~\ref{sec:games}. 
We introduce tree decomposition for scenarios in Sect.~\ref{sec:tdecs},
and we link tree decompositions to monotone winning strategies.
We introduce brambles and we show that they provide a
strategy for the robber to escape.
In Sect.~\ref{sec:monotonicity} we introduce weakly submodular
scenarios and search trees. We prove monotonicity, linking
brambles, search trees, tree decompositions and winning strategies
together.
Sect.~\ref{sec:bdecs} introduces branch decompositions
for scenarios and shows how they relate to tree decompositions
of scenarios.
Sect.~\ref{sec:applications} contains applications to matroid
tree-width, to graph tree-width and to branch-width of connectivity 
functions, yielding monotone games for each of the invariants.
As an aside, Sect.~\ref{subsec:twundaside} contains
proof showing that
the matroid tree-width of a graphic matroid is not larger
than the (traditional) tree-width of the corresponding
graph. Our proof is much shorter than the original
proof, and it is purely graph theoretic, avoiding the geometric
argument in the original proof~\cite{HlinenyW06}.
Finally, we close with a conclusion in Sect.~\ref{sec:conclusion}.


\section{Scenarios and Games}\label{sec:games}

For an integer $n\geq 1$ let $[n]:=\{1,\ldots,n\}$. 
A \emph{partition} of a set
of $A$ is a set $P=\{A_1,\ldots, A_{d}\}$, consisting
of pairwise disjoint subsets $A_i\subseteq A$
such that $A=A_1\,\dot\cup\, \ldots\,\dot\cup\,A_{d}$.
We allow the sets $A_i$ to be empty.
Let $P_1=\{A_1,\ldots, A_{d}\}$ and
$P_2=\{B_1, \ldots, B_{\ell}\}$ be two partitions
of $A$ into sets $A_i\subseteq A$ and
$B_j\subseteq A$, respectively. 
We say that $P_1$ is \emph{coarser} than $P_2$, $P_1\geq P_2$, 
if every set in $P_1$ is a union of some sets of $P_2$, i.e. for all 
$i\in[d]$ there exist $i_1,\ldots, i_n\in[\ell]$
such that $A_i=B_{i_1}\,\dot\cup\, \ldots\,\dot\cup\,B_{i_n}$
(we also say that $P_2$ is \emph{finer} than $P_1$ and write $P_2\leq P_1$).

The \emph{common coarsening}
of the two partitions is the partition $P_1\vee P_2$ of $A$ into
subsets $X\subseteq A$ that can be written as a union of $A_i$s as well
as a union of $B_i$s, i.e.\ $X=A_{i_1}\,\dot\cup\, \ldots\,\dot\cup\,A_{i_n}$
for some $i_1,\ldots, i_n\in[d]$, and
$X=B_{i_1}\,\dot\cup\, \ldots\,\dot\cup\,B_{i_m}$
for some $i_1,\ldots, i_m\in[\ell]$.
By $\Part(A)$ we denote the collection of all partitions of $A$. Note that
$(\Part(A),\leq)$ is a lattice.
By $2^A$ we denote the set of all subsets of $A$, and for a subset
$X\subseteq A$, we let $X^c:=A\setminus X$ denote the complement of $X$ in $A$.

\begin{dfn} Let $A$ be a finite set. A \emph{scenario} on $A$
	is a pair $(\mP,\mS)$, where $\mP\subseteq\Part(A)$ and
	$\mS\subseteq 2^A$ satisfy
	\begin{description}
		\item[(SC1)] $\mP$ is closed under coarser partitions,
		\item[(SC2)] If $X\subseteq S$ for some $S\in\mS$ and
			there is a partition $P\in\mP$ with $X\in P$, then
			$X\in\mS$,
		\item[(SC3)] Every set $S\in\mS$ satisfies 
			$\{S,S^c\}\in\mP$.
	\end{description}
\end{dfn}
Note that $\mP=\emptyset$ implies $\mS=\emptyset$, and that
$\mP\neq\emptyset$ implies
$\{A\}\in\mP$.
Intuitively, the set $\mP$ is the set of `feasible' partitions, 
and $\mS$ contains the `simple' subsets
of $A$ -- the subsets that are `well understood'.
By (SC1), making partition coarser is `feasible'.
Condition (SC2) says that if $S$ is a subset of a `simple'
set, and if we can border $X$ with a
feasible partition, then $X$ is `simple' as well.
According to the last condition, 
`simple' subsets should have a `feasible' border.

\begin{prov}
	Throughout the whole paper, $A$ denotes a nonempty, 
	finite set,
	$\mP$ denotes a set of partitions of $A$,
	and $\mS$ denotes a collection of subsets of $A$.
\end{prov}

\paragraph{The captain and robber game}

Let $(\mP,\mS)$ be a scenario on $A$. The \emph{captain and robber game} on $(\mP,\mS)$ is a two player game, 
where one player controls the captain and the other player controls the
robber. The robber moves on elements of $A$, running within 
certain subsets of $A$. 
The captain, sitting in her office, lets her assistants build 
barriers in order to limit the way the robber can move. 
Such a barrier must be `feasible', 
so her choice is limited to a set $\mP$ of 
`allowed' barriers. The captain's goal
is to limit the robber to a set $S\in \mS$ (intuitively, the sets in $\mS$ 
are well-known areas, where it is easy to catch the robber).
The robber's goal is to avoid being cornered in any set of $\mS$.

More precisely, in the beginning of a play the captain 
has not blocked anything, 
i.e. she chooses the trivial partition $\{A\}$ of $A$,
and the robber moves to a arbitrary element of $A$. If $\{A\}\notin\mP$,
then the robber wins. Otherwise
we have $\{A\}\in\mP$, and hence the 
partition $\{A\}$ is an allowed choice and the game continues. 
Now suppose the game is in position $(P,r)$, where $P\in \mP$ is the partition
chosen by the captain and the robber stands on $r\in A$. 
Then $r\in X\in P$ for some set $X$ in the partition $P$. (The set $X$ is
called the \emph{robber space}.) 
Now the captain chooses a new partition $P'\in \mP$. The clever robber finds out
which partition $P'$ she chooses. Now let $Y\in P\vee P'$ be the subset satisfying
$X\subseteq Y$. While the barriers are moved from
$P$ to $P'$, the captain only blocks 
$P\vee P'$, and
the robber can move within the borders of 
$P\vee P'$ to a (possibly) new element
$r'\in Y$. Then the new 
barrier is built and the robber
is in the set $X'\subseteq Y$ with $r'\in X'\in P'$. If $X'\in\mS$, 
then the captain wins. Otherwise
the play continues. The captain wins if in some step of the play 
she catches the robber within
a set of $\mS$. Otherwise the robber wins.

The captain has a \emph{winning strategy} on $(\mP,\mS)$ 
if she can assure capturing the robber 
independently of the way the robber moves. Winning strategies for
the robber are defined analogously.

The \emph{monotone captain and robber game} on $(\mP,\mS)$
is defined like the captain and robber game on $(\mP,\mS)$,
with the additional restriction that all the captain's moves be \emph{monotone}:
Suppose that the play is in position $(P,r)$, where $P\in \mP$ is the partition
chosen by the captain and the robber stands on $r\in X\in P$. Now the captain
may only choose a partition $P'\in\mP$ that refines $X$, 
i.e.~$P'$ contains subsets $X'_1,\ldots X'_n$
such that $X=X'_1\cup\ldots\cup X'_n$. 
Note that this restriction assures that
after moving to $P'$ the robber space either stays the same or decreases.

Winning strategies for the captain and for the robber in 
the monotone captain and robber game are defined
as usual.

\begin{rem}\label{rem:game-mongame}
Let $(\mP,\mS)$ be a scenario on $A$.  

If the captain has a winning strategy in the monotone captain and robber game 
on $(\mP,\mS)$, then she has a winning strategy in the (non-monotone)
captain and robber game on $(\mP,\mS)$.\qed
\end{rem}

Note that if $\bigcup \mS \subsetneqq A$, then the robber can win by
always staying on an element $r\in A\setminus \bigcup \mS$.
 
\begin{dfn}
A scenario $(\mP,\mS)$ is \emph{monotone},
if the captain has a winning strategy in the 
captain and robber game on $(\mP,\mS)$
if and only if the captain has a winning strategy in
the monotone captain and robber game on $(\mP,\mS)$. 
\end{dfn}
In Sect.~\ref{sec:monotonicity} we exhibit classes of monotone scenarios.


\section{Tree Decompositions and Brambles for Scenarios}\label{sec:tdecs}

Graphs are finite, simple and undirected, unless stated otherwise.
For a graph $G$ we denote the vertex set by $V(G)$ and the edge set
by $E(G)$. For a vertex $u\in V(G)$ we let 
$N_G(u):=\big\{v\in V(G)\mid \{u,v\}\in E(G)\big\}$ denote the
set of \emph{neighbours} of $u$ in $G$ (we omit the subscript $G$ if
it is clear from the context). A tree is a nonempty, connected, acyclic graph.
(For the basic notions of graph theory see \cite{die05}).
For a tree $T$ let $L(T)$ denote the set of leaves of $T$, i.e.\ the
nodes of degree at most one. We call the nodes of $V(T)\setminus L(T)$
\emph{internal} nodes. For $t\in V(T)$
let $T^{-t}$ denote the set of connected components of $T\setminus t$.
Let $(\mP,\mS)$ be a scenario on $A$, and let $\tau\colon L(T)\to\mS$ 
be a mapping from the set of leaves of $T$ to $\mS$. 
For an internal node $t$ of $T$ we let
$P_t:=\{\bigcup\tau\big(L(T)\cap V(T')\big)\mid T'\in T^{-t}\}$. 

A \emph{tree decomposition} for a scenario $(\mP,\mS)$ is a pair $(T,\tau)$,
where $T$ is a tree and $\tau\colon L(T)\to\mS$ is a mapping from
the set of leaves of $T$ to $\mS$, such that
\begin{description}
	\item[(TD1)] the image $\tau\big(L(T)\big)\subseteq\mS$ 
		is a partition of $A$, and 	
	\item[(TD2)] all internal nodes $t\in V(T)\setminus L(T)$ satisfy
		$P_t\in \mP$.
\end{description}

Note that we do not require the partition $\tau\big(L(T)\big)$ to be in $\mP$:
Assume there exists a partition $P=\{S_1,\ldots,S_n\}\in\mP$
with $S_i\in\mS$ for all $i\in[n]$. Then the $n$-star $T_n$, i.e.\ the 
tree $T_n$ consisting of one node $s$ of degree $n$ and $n$ leaves $t_1,\ldots,t_n$,
together with the mapping $\tau(t_i):=S_i$, is a tree decomposition
for $(\mP,\mS)$. 
Note that if $\bigcup\mS\neq A$, then $(\mP,\mS)$ has no tree decomposition.

\begin{theo}\label{theo:tdec-game}
Let $(\mP,\mS)$ be a scenario on $A$.  

The pair $(\mP,\mS)$ has a tree decomposition if and only if
the captain has a winning strategy in the monotone captain and robber game
on $(\mP,\mS)$.
\end{theo}

A proof sketch of Theorem \ref{theo:tdec-game} can be found in the appendix.

\medskip 
A \emph{bramble} for $\mP$ is a nonempty collection $\mB$
of nonempty, pairwise intersecting subsets of $A$, such that
every partition $P\in\mP$ satisfies $P\cap\mB\neq\emptyset$.
A bramble $\mB$ for $\mP$ 
\emph{avoids} $\mS$, if $\mB\cap\mS=\emptyset$.

\begin{lem}\label{lem:game-bramble}
	Let $(\mP,\mS)$ be a scenario on $A$.

	If $\mP$ has a bramble avoiding $\mS$, then the robber has a winning
	strategy in the captain and robber game on $(\mP,\mS)$.
\end{lem} 

\proof
The robber can escape: 
Whenever the captain chooses a partition $P\in \mP$, the robber moves
to the set $X\in\mP\cap\mB$. This is always possible since
any two sets in $\mB$ have a nonempty intersection.
\qed


\section{Monotonicity of Weakly Submodular Scenarios}\label{sec:monotonicity}

In this section we prove monotonicity for a
class of scenarios with \emph{weakly
submodular} sets of partitions. The proof uses
the notion of search trees for scenarios.
For our proof of monotonicity we generalise methods 
of \cite{LyaudetMT09} to scenarios
(The paper \cite{LyaudetMT09} simplifies and slightly generalises
the ideas of \cite{ammanith08}).

Let $P=\{X_1,\ldots,X_d\}$ be a partition of $A$ and let $F\subseteq A$. 
For $i\in[d]$ let $P_{X_i\to F}$ be the partition
$P_{X_i\to F}:=\{X_1\cap F^c,\ldots,X_{i-1}\cap F^c,X_i\cup F,
X_{i+1}\cap F^c,\ldots,X_d\cap F^c\}.$

The following notion of weak submodularity is crucial to the proof of
monotonicity: In Lemma~\ref{lem:make-exact},
we need to rearrange partitions induced by tree labelings.
Weakly submodular sets of partitions allow for the necessary
rearrangements.

We say that a set $\mP$ of partitions of $A$ is \emph{weakly
submodular}\footnote{
This is a translation from the notion of
\emph{weakly submodular partition function} of 
\cite{LyaudetMT09} into the context
of scenarios.}, if for any pair of partitions $P,Q\in\mP$
and any pair of sets $X\in P$ and $Y\in Q$ with 
$A\setminus (X\cup Y)\neq\emptyset$ there exists a nonempty set 
$F\subseteq A\setminus (X\cup Y)$ such that $P_{X\to F}\in\mP$, or
$Q_{Y\to F}\in\mP$.

\begin{dfn} Let $A$ be a finite set.
A scenario $(\mP,\mS)$ on $A$ is \emph{weakly submodular}, if
$\mP$ is weakly submodular.
\end{dfn}

The two following
propositions present examples of weakly submodular sets of partitions.
Let $G$ be a graph and let
$P=\{X_1,\ldots,X_d\}$ be a partition of $E(G)$.
Define
\[\partial(P):=\{v\in V(G)\mid\exists\; e_i,e_j\in E(G) \text{ with }
v\in e_i\cap e_j,\; e_i\in X_i,\;e_j\in X_j, \text{ and } i\neq j\}.\] 

For an integer $k\geq 1$ we let 
$\Ptwk:=\big\{P\in\Part(E(G))\bigmid\left|\partial(P)\right| \leq k\big\}$.
It is straightforward to check that the following holds.

\begin{prop}\label{prop:partial}
Let $G$ be a graph and let $k\geq 1$ be an integer.
Then $\Ptwk$ is a weakly submodular set of partitions, and $\Ptwk$ is closed
under coarser partitions.  \qed
\end{prop}

More generally, we consider sets of partitions arising from
connectivity functions. Connectivity functions are strongly related with 
matroids, and they arise in many different contexts 
(see i.e.~\cite{Oxley92}). Let $f$ be an integer valued
function $f\colon 2^A\to\mathbb Z$. The function $f$ is a
\emph{connectivity function on} $A$, if

\begin{itemize}
	\item any subset $X\subseteq A$ satisfies $f(X)=f(X^c)$ (\emph{symmetry}),
	\item any two subsets $X,Y\subseteq A$ satisfy 
		$f(X)+f(Y)\geq f(X\cup Y)+f(X\cap Y)$ (\emph{submodularity}).
\end{itemize}

\begin{ex}\label{ex:delta}
Let $G$ be a graph. The function
$\delta\colon 2^{E(G)}\to \mathbb Z$, given by 
$\delta(X):=\left|\partial(\{X,X^ c\})\right|$ for $X\subseteq E(G)$,
is a connectivity function on $E(G)$.
\end{ex}

For a connectivity function $f$ and $k\in\mathbb Z$,
we let 
$\Pfk:=\{P\in\Part(A)\mid \sum_{X\in P}f(X)\leq k\}.$

\begin{prop}\label{prop:submod-partitions}
Let $f$ be a connectivity function on $A$ and let $k$ be an integer.
Then $\Pfk$ is a weakly submodular set of partitions, and $\Pfk$ is closed
under coarser partitions.  
\end{prop}

\proof	
Submodularity of $f$ implies that $\Pfk$ is closed
under coarser partitions. It is straightforward to check
that  $\Pfk$ is a weakly submodular (cf.~\cite[Sect.~6]{ammanith08}).
\qed

\medskip\noindent
A \emph{bidirected tree} is 
obtained from an undirected tree with at least one edge
by replacing every edge by two edges directed in opposite directions. 
Directed edges are also called \emph{arcs}.
\emph{Neighbours} in a bidirected tree are neighbours in the
underlying undirected tree. Let $T$ be a tree and let 
$l\colon E(T)\to 2^A$. For an internal node $t$ of $T$
with neighbours $t_1,\ldots,t_n$ we let 
$\pi_t:=\big\{l(t,t_i)\bigmid i\in[n]\big\}$. 

A \emph{search tree} for $A$ is a pair $(T,l)$, where $T$ is a 
bidirected tree, and 
$l\colon E(T)\to 2^A$
is a labeling function such that
\begin{description}
	\item[(ST1)]  for every internal node $t\in V(T)\setminus L(T)$ 
		the set $\pi_t$ 
		is a partition of $A$, and 
	\item[(ST2)] the two labels of every $2$-cycle are disjoint, i.e.
		$l(s,t)\cap l(t,s)=\emptyset$ for all $(s,t)\in E(T)$.
\end{description}
	
A $2$-cycle $st$ of $T$ is \emph{exact}, if $l(s,t)\cup l(t,s)=A$.
A search tree $(T,l)$ is \emph{exact}, if all its $2$-cycles are 
exact.
The following is proved in \cite{ammanith08}.
\begin{fact}\label{fact:exact-partition}
	In an exact search tree $(T,l)$ for $A$, the labels of the arcs entering 
	the leaves form a partition of $A$. \qed
\end{fact}

A label of an arc leaving a leaf is called a \emph{leaf label}. 
Note that
in an exact search tree, a leaf label 
other than $A$ cannot appear 
twice. 

A search tree $(T,l)$ for $A$ is a \emph{search tree for} $\mP$, 
where in addition all internal nodes of $T$ satisfy 
$\pi_t\in\mP$.
We extend this definition to scenarios. 
A search tree $(T,l)$ for $\mP$ is a \emph{search tree for} $(\mP,\mS)$, 
if, in addition, every $(s,t)\in E(T)$ with $t\in L(T)$ satisfies 
$l(s,t)\in\mS$.
A search tree $(T,l)$ is \emph{compatible} with a set $\mF\subseteq 2^A$, if
every leaf label contains an element of $\mF$ as a subset.
The following Lemma is proved in the appendix.

\begin{lem}\label{lem:make-exact}
	Let $A$ be a finite set, let $(\mP,\mS)$ be a weakly submodular scenario for $A$,
	and let $\mF\subseteq 2^A$. 
	
	If $(\mP,\mS)$ has a search tree compatible with $\mF$ 
	having at least one internal node, 
	then $(\mP,\mS)$ has an \textup{exact} search tree  
	compatible with $\mF$.
\end{lem}

Note that if $(T,l)$ is an exact search tree for $(\mP,\mS)$, then by
Fact \ref{fact:exact-partition}, the labels entering the leaves of $T$ form 
a partition of $A$ into subsets from $\mS$. This is the link between search trees
and tree decompositions. The proof of the following Theorem is given
in the appendix.

\begin{theo}\label{theo:seachtree-tdec}
	Let $A$ be a finite set and let $(\mP,\mS)$ be a scenario for 
	$A$. 
	If the pair	$(\mP,\mS)$ has an exact search tree, then it has
	a tree decomposition.
\end{theo} 

For $\mS\subseteq 2^A$, and
$\mS$-\emph{bias} in $A$ is a nonempty set $\mB\subseteq 2^A$ 
	of nonempty subsets of $A$ satisfying
\begin{itemize}
	\item for every $S\in\mS$ there is a set $X\in\mB$ such that
		$S\cap X=\emptyset$,
	\item $\mB\cap\mS=\emptyset$. 
\end{itemize}

For example, let $\left|A\right|\geq 2$ and let $\bigcup\mS=A$.
Suppose every partition $P\in\Part(A)$ with $\left|P\right|\leq 2$
satisfies $P\not\subseteq \mS$. Then 
$\{S^c\mid S\in\mS\}$ is an $\mS$-bias in $A$.
We remark that for $\mS_1:=\{\{a\}\mid a\in A\}$, 
an $\mS_1$-bias is a \emph{bias} as defined in \cite{ammanith08}. 
Moreover, brambles avoiding $\mS_1$ are precisely the \emph{non-principal}
brambles from \cite{ammanith08}.

The following theorem generalises Theorem~4 of \cite{ammanith08}  
and, with it, Theorem~3.4 of \cite{GMX}.

\begin{theo}\label{theo:duality}
	Let $A$ be a finite set and let $(\mP,\mS)$ be a weakly 
	submodular scenario for $A$, satisfying $\bigcup\mS=A$. 
		
	If the set of partitions $\mP$ has no bramble avoiding $\mS$, 
	then $(\mP,\mS)$ has an exact search tree. 
\end{theo}

\proof 
If $\mP=\emptyset$, then $\mS=\emptyset$ and every bramble avoids
$\mS$. Suppose now that $\mP\neq\emptyset$. 
If $A\in\mS$, then there is no bramble avoiding
$\mS$ and we obtain an exact search tree for
$(\mP,\mS)$ by taking two nodes $s,t$ with labels $l(s,t)=A$
and $l(t,s)=\emptyset$. 
If there is a bipartition $\{X, X^c\}\in\mP$ satisfying
$\{X, X^c\}\subseteq \mS$, then the two node tree
with labels $X$ and $X^c$ is an exact search tree for $(\mP,\mS)$.

For the rest of the proof, assume that $A\notin\mS$ and
that all bipartitions $P\in\mP$ satisfy $P\not\subseteq\mS$.
It is easy to check that in this case, the set
$\mB_c:=\{S^c\mid S\in\mS\}$ is an $\mS$-bias
in $A$.

\medskip\noindent
\emph{Claim.} For every $\mS$-bias $\mB$ in $A$ there is a
search tree for $\mP$ compatible with $\mB$. 

\medskip\noindent
\emph{Proof of the Claim.}
Towards a contradiction,
assume that there is no search tree for $\mP$ compatible with 
$\mB$. Choose $\mB$ of maximum cardinality with this property.

First assume that for every partition $P\in\mP$ there
exists a set $X_P\in\mP\cap\mB$. Since $\mB$ is an
$\mS$-bias, we have $X_P\notin\mS$. 
Since $\mP$ has no bramble avoiding $\mS$, 
there must be two sets $X,Y\in\mB$ with $X\cap Y=\emptyset$.
But then the $2$-cycle labeled $X$ and $Y$ is a
search tree for $\mP$ compatible with $\mB$, a contradiction.

Secondly, assume there is a partition $P=\{X_1,\ldots X_n\}\in\mP$
such that $P\cap\mB=\emptyset$. 

\smallskip\noindent
$(1)$ For every $i\in[n]$ satisfying $X_i\notin\mS$ 
there exists a search tree $(T_i,l_i)$
for $\mP$ that has exactly one leaf label containing $X_i$ as a
subset, and all other leaf labels contain an element of
$\mB$ as a subset.

\smallskip\noindent
\emph{Proof of} $(1)$. Let $i\in[n]$ satisfy
$X_i\notin\mS$. Choose a superset $X'_i\supseteq X_i$ 
of maximum cardinality such that $X_i'\notin\mB$. Then $\mB\cup\{X_i'\}$ is an
$\mS$-bias. By maximality of $\mB$, there is a search tree for $\mP$
compatible with $\mB\cup\{X_i'\}$, and by Lemma~\ref{lem:make-exact}
there is an exact search tree $(T_i,l_i)$ for $\mB\cup\{X_i'\}$.
If $(T_i,l_i)$ also is a search tree compatible $\mB$ we are done.
Otherwise there exists a leaf $t_i$ with a leaf label containing
$X_i'$ and containing no other element of $\mB$ as a subset. 
By maximality of $X_i'$, the leaf label is exactly 
$X_i'$. Note that $X_i'\neq A$, since $\mB$ contains at least one
nonempty set which is not contained in $X_i'$. Hence
by Fact~\ref{fact:exact-partition}
there is exactly one leaf label $X_i'$. This proves $(1)$.

With $(1)$ we complete the proof of the claim.
For every $i\in[n]$ with $X_i\notin\mS$ 
let $t_i\in L(T_i)$ be the leaf
with the label containing $X_i$ as a subset. 
We glue the trees $T_i$ together by identifying all 
the nodes $t_i$ into a new node
$t$. Then the neighbour $s_i$ of $t_i$ in $T_i$
the becomes a neighbour of $t$ in the new tree $T$, and we label
$(t,s_i)$ by $X_i$ and keep all other labels as in $(T_i,l_i)$.
For every $j\in[n]$ satisfying $X_j\in\mS$ we add a new node $t_j$
to $T$ via a $2$-arc
$tt_j$, labeling $(t,t_j)$ by $X_j$ and $(t_j,t)$ by $X_j^c$.
It is easy to see that this gives us a search tree for  
$\mP$ compatible with $\mB$.\\ 
This proves the claim.

\medskip
Now we choose a search tree $(T,l)$ 
for $\mP$ compatible with $\mB_c$, which exists according to the claim.
First suppose $T$ consists of a single edge $V(T)=\{s,t\}$ 
with $X:=l(s,t)$ and $Y:=l(t,s)$. 
Then there exist sets $X_0\in\mB$ and $Y_0\in\mB_c$
with $X_0\subseteq X$ and $Y_0\subseteq Y$. By the definition
of $\mB_c$ we have
$\{X_0^c,Y_0^c\}\subseteq\mS$, and by (SC3)   
we have $\{X_0,X_0^c\}\in\mP$. 
We now replace $Y$ by the new label $X^c$. Then the $2$-cycle
is exact, it remains compatible with $\mF$, 
and we have $X^c\in\mS$. 
We claim that $X\in\mS$:  
This follows from $X\subseteq Y^c\in\mS$ and $\{X,X^c\}\in\mP$
using (SC2). Hence we have found an exact search tree for 
$(\mP,\mS)$.

Secondly, suppose $T$ has at least one internal node. 
Let $s\in L(T)$ and let $t\in V(T)$ be the neighbour of $t$.
Then $l(s,t)$ is a leaf label, and hence there is a subset
$X_{st}\subseteq l(s,t)$ with $X_{st}\in\mB_c$. Define a new
labeling of $T$ by letting
$l'(s,t):=X_st$ and $l'(t,s):=X_{st}^c$ for all arcs $(s,t)$
where $s\in L(T)$, and letting $l'(e)=l(e)$
for all other arcs of $T$.
Then the labels of the arcs entering a leaf are in $\mB$.
By Property (SC2), the pair $(T,l')$ is a search tree for $(\mP,\mS)$.
Now we apply Lemma~\ref{lem:make-exact} and we obtain an exact
search tree for $(\mP,\mS)$. 
\qed

\medskip 
Note that the condition $\bigcup\mS=A$ is necessary: Otherwise,
we can choose an element $a\in A\setminus\bigcup\mS$ and define
the bramble $\mB:=\{X\subseteq A\mid a\in X,\; X\notin\mS\}$.
Then $\mB$ avoids $\mS$, but $(\mP,\mS)$ has no tree decomposition
and hence by Theorem \ref{theo:seachtree-tdec} 
it has no search tree.

Combining Theorem~\ref{theo:seachtree-tdec}, Theorm~\ref{theo:tdec-game},
Theorem~\ref{theo:duality},
Lemma~\ref{lem:game-bramble}, and Remark~\ref{rem:game-mongame}
we obtain the following Corollary.

\begin{cor}[Characterising Tree Decomposable Scenarios]\label{cor:biglist}
Let $A$ be a finite set and let 
$(\mP,\mS)$ be a weakly submodular scenario on $A$, 
satisfying $\bigcup\mS=A$. 
	Then the following statements are equivalent.
	\begin{enumerate}
		\item The pair $(\mP,\mS)$ has an exact search tree.
		\item The pair $(\mP,\mS)$ has a tree decomposition.
		\item The captain has a winning strategy in
			the monotone captain and robber game on $(\mP,\mS)$.
		\item The captain has a winning strategy in
			the captain and robber game on $(\mP,\mS)$.
		\item The set of partitions $\mP$ has no bramble avoiding $\mS$.
	\end{enumerate}
\end{cor}

\begin{cor}\label{cor:ws-scenarios} Let $A$ be a finite set.
All weakly submodular scenarios $(\mP,\mS)$ on $A$ satisfying $\bigcup\mS=A$ are monotone. 
\end{cor}

\section{Branch Decompositions for Scenarios}\label{sec:bdecs} 

Branch-width of graphs is closely related to
tree-width of graphs \cite{GMX}. We generalise
branch decompositions to our setting, maintaining the
close relation in a natural way.
Let $T$ be a tree and let $\beta\colon L(T)\to A$ be a mapping.
For $e\in E(T)$, let $T_1,T_2$ denote the two connected
components of $T-e$, and let $P_e$ denote the pair
$P_e:=\big\{\beta\big(L(T)\cap V(T_1)\big),
\beta\big(L(T)\cap V(T_2)\big)\big\}$
of subsets $A$.

A tree $T$ is cubic, if every internal node of $T$ has degree $3$.
Let $(\mP,\mS)$ be a scenario.
A \emph{branch decomposition} for $(\mP,\mS)$ is a
pair $(T,\beta)$ where $T$ is a cubic tree, and
$\beta\colon L(T)\to \mS$ is a mapping, such that
\begin{description}
	\item[(BD1)] The image $\beta\big(L(T)\big)\subseteq\mS$ 
		is a partition of $A$, and
	\item[(BD2)] every edge $e\in E(T)$ satisfies
		$P_e\in\mP$.
\end{description}

There is a close link between branch decompositions
and tree decompositions. The following theorem is proved in the appendix.

\begin{theo}\label{theo:tdec-bdec}
	Let $(\mP,\mS)$ be a scenario on $A$.
If $(\mP,\mS)$ has a tree decomposition, then
$(\mP,\mS)$ has a branch decomposition. 
\end{theo}

Let $\mP\subseteq \Part(A)$. We define the set of partitions
$\mP^3\subseteq\Part(A)$ by\\
$
\mP^3:=\big\{\{X,Y,Z\}\bigmid 
\big\{\{X,X^c\},\{Y,Y^c\},\{Z,Z^c\}\big\}\subseteq \mP,\;
X\dot\cup Y\dot\cup Z=A \big\}.
$

\begin{rem}\label{rem:bdec-tdec}
Let $(\mP,\mS)$ be a scenario on $A$.

If $(T,\beta)$ is a branch decomposition for $(\mP,\mS)$, then 
$(T,\beta)$ is a tree decomposition for $(\mP^3,\mS)$.
\end{rem}

\proof
Every branch decomposition $(T,\beta)$ for $\mP$ is
a tree decomposition for $(\mP^3,\mS)$.
\qed
\section{Applications to Width Parameters}\label{sec:applications} 

\subsection{Branch-width of Connectivity Functions}\label{subsec:connectivity} 

Given a connectivity function $f$, we
approximate the branch-width of $f$ by the captain and robber 
game. In particular, we obtain a game equivalent to rank-width of 
graphs~\cite{oumsey05}, and games equivalent to
$\gf$-rank-width and bi-rank-width of directed graphs \cite{kante07}.
All these scenarios are monotone.
Let $A$ be a nonempty, finite set, let $f:2^A\to \mathbb Z$ be a 
connectivity function and let $k$ be an integer.
Recall that $\Pfk=\{P\in\Part(A)\mid\sum_{X\in P}f(X) \leq k\}$
(cf.~Proposition~\ref{prop:submod-partitions}).

\begin{theo}\label{theo:bw}
Let $A$ be a finite set and let $f:2^A\to \mathbb Z$ be a 
connectivity function, and let 
$\mS\subseteq \Part(A)$ be closed under subsets
(i.e.~if $S'\subseteq S\in\mS$, then $S'\in\mS$). 
Let $k$ be an integer satisfying 
$k\geq \max\{f(S)\mid S\in\mS\}$. Then
\begin{enumerate}
	\item $(\mP_f^k,\mS)$ is a weakly submodular scenario on $A$,
	\item $(\mP_f^k,\mS)$ satisfies 
		Corollary~\ref{cor:biglist} (Characterising tree decomposable scenarios),
	\item In particular, the scenario $(\mP_f^k,\mS)$ is monotone.
\end{enumerate}
\end{theo}

\proof
1: By Proposition~\ref{prop:submod-partitions}, the set 
$\mP_f^k$ is closed under coarser partitions, and it is weakly submodular.
In particular, $(\mP_f^k,\mS)$ satisfies (SC1). 
By the choice of $k$, it satisfies (SC3) as well.
Since $\mS$ is closed under subsets, it satisfies (SC2).
Hence $(\mP_f^k,\mS)$ is a weakly submodular scenario on $A$.
Statements 2 and 3 follow from 1, together with Corollary~\ref{cor:biglist}.
\qed

\medskip 
Let us apply Theorem \ref{theo:bw} to a graph
$G$ by letting $A=E(G)$ and $f:=\delta$ (cf.~Example \ref{ex:delta}).
Now we choose our favorite class $\mathcal C$ of graphs that is closed under
taking subgraphs (planar, $H$-minor free, etc.), and let 
$\mS:=\{S\subseteq E(G)\mid G[S]\in\mathcal C\}$.
Choosing $k$ as in the theorem, we obtain a weakly submodular
scenario $(\mP_{\delta}^k,\mS)$ satisfying Corollary~\ref{cor:biglist}
(Characterising Tree Decomposable Scenarios).

For a connectivity function $f$ on $A$ and an integer $k$,
let $\mathcal Q_f:=\big\{\{X,X^c\}\in\Part(A)\bigmid f(X)\leq k\big\}$.
Let $\Ssing:=\big\{\{a\}\mid a \in A\big\}\cup\{\emptyset\}$,
and assume that $k\geq \max\big\{f(\{a\})\mid a\in A\big\}$. 
Then, by Theorem \ref{theo:bw}, $(\mathcal Q_f, \Ssing)$ is a scenario. 
We say that $f$ has \emph{branch-width} at most $k$, $\bw(f)\leq k$, if
the scenario $(\mathcal Q_f, \Ssing)$ has a branch decomposition.
This is equivalent to the conventional definition
of branch-width (see i.e.~\cite{HlinenyO07}).
Note that $\mathcal Q_f^k\subseteq\mP_f^k$, and
that $(\mathcal Q_f, \Ssing)$ has a branch decomposition
if and only if $(\mP_f^k,\Ssing)$ has a branch decomposition.
Analogously, the following definition extends the definition of tree-width
of graphs to tree-width of submodular functions. 

\begin{dfn}
Let $f$ be a connectivity function on $A$, and 
let $\Ssing:=\big\{\{a\}\mid a \in A\big\}\cup\{\emptyset\}$.
Let $k$ be an integer satisfying 
$k\geq \max\big\{f(\{a\})\mid a\in A\big\}$. 

We say that $f$ has \emph{tree-width} at most $k$, $\tw(f)\leq k$, if
the scenario $(\mathcal P_f^k, \Ssing)$ has a tree decomposition.
\end{dfn}

\begin{cor} 
\begin{enumerate}
	\item	If $(\mP_f^k,\Ssing)$ has a tree decomposition, then
		$(\mP_f^k,\Ssing)$ has a branch decomposition.
	\item If $(\mP_f^k,\Ssing)$ has a branch decomposition,
		then $((\mP_f^k)^3,\Ssing)$ has a tree decomposition.
	\item $\bw(f)\leq\tw(f)\leq 3\cdot\bw(f)$.
\end{enumerate}
\end{cor}

\proof
1 and 2 follow from Theorem~\ref{theo:tdec-bdec} and Remark~\ref{rem:bdec-tdec}.
3 follows from 1 and 2, using submodularity of $f$.
\qed

\medskip
Hence branch-width and tree-width of a submodular function $f$
are within a factor of three of each other, and the tree-width of
$f$ can be characterised by a monotone game. In particular, this applies 
to rank-width of 
graphs, to $\gf$-rank-width and to bi-rank-width of directed graphs.

\begin{cor}
Rank-width and carving-width of graphs, and both
$\gf$-rank-width and bi-rank-width of directed graphs
have factor $3$ approximations by monotone 
games, that can also be characterised by tree decompositions.
\end{cor}


\subsection{Tree-width of Matroids}\label{subsec:mtw}

Matroid tree-width was introduced by 
Hlinen{\'y} and Whittle in \cite{HlinenyW06}.
In this section, we present the scenario for the game
characterising matroid tree-width. This scenario is monotone. Moreover, 
we include a short proof showing that
the matroid tree-width of a graphic matroid is not larger
than the (traditional) tree-width of the corresponding
graph.

Throughout this section, let $M$ be a matroid with 
nonempty ground set $E=E(M)$, and let $\r$ be the \emph{rank function} of 
$M$ (see \cite{Oxley92} for an introduction into matroid theory). 

A \emph{tree decomposition} for $M$ is a pair
$(T,\iota)$ where $T$ is a tree and $\iota\colon E\to V(T)$
is an arbitrary mapping.
For a node $x\in V(T)$ let $T^x_1,\ldots,T^x_d$ denote the 
connected components of $T-x$, and let
$F^x_i:=\iota^{-1}\big(V(T^x_i)\big)$ (hence $F^x_i\subseteq E$).
The \emph{node-width} of $x$ is defined by
\[
	\nw(x)=\sum_{i=1}^d\r(E\setminus F^x_i)-(d-1)\cdot\r(M).
\]
The \emph{width} of the decomposition is the maximum width 
of the nodes of $T$, and the smallest width over all tree 
decompositions of $M$ is the \emph{(matroid) tree-width} of $M$,
denoted by $\mtw(M)$. (The width of an empty tree is $0$.)

For a better understanding of node-width, we give two equivalent
formulations. Let $\lambda_M\colon 2^E\to\mathbb N$ with
$\lambda_M(X)=\r(X)+\r(E-X)-\r(M)$, denote the \emph{connectivity 
function} on $M$.

\begin{rem}
	Let $(T,\iota)$ be a tree decomposition of a matroid $M$
	and let $x\in V(T)$. Then
	\[
		\nw(x)=\r(M)-\sum_{i=1}^d\big[\r(M)- \r(E-F^x_i)\big]
		=\r(M)-\sum_{i=1}^d\big[\r(F^x_i)-\lambda_M(F^x_i)\big].
	\]
\end{rem}

For a set $F\subseteq E$ the \emph{rank defect} (cf.\ \cite{HlinenyW06}) 
of $F$ is given by
$r(M)-\r(E-M)$. So for small node width, the second term intuitively says
that we want to maximize the rank defect on the branches of $T-x$.
Similarly, the third term says that we want to maximize the rank of each
branch of $T-x$ using small cuts.

Notice that we do not require $\iota$ to be surjective.
We can actually restrict the image $\iota(E(M))$ to the leaves of the 
decomposition tree (we prove this in the appendix): 

\begin{lem}\label{lem:move-to-leaves}
Let $M$ be a matroid with $\mtw(M)\leq k$. There is
a tree decomposition $(T,\iota)$ for $M$ of width at most $k$
satisfying $\iota(E(M))\subseteq L(T)$.
\end{lem}

Let $P=\{X_1,\ldots, X_d\}$ be a partition of $E:=E(M)$.
The \emph{width} of
$P$ is defined as
$
	\w(P)=\sum_{i=1}^d\r(E\setminus X_i)-(d-1)\cdot\r(M).
$
Let $(T,\iota)$ be a tree decomposition for a matroid $M$
where $\iota(E(M))\subseteq L(T)$. 
For $x\in V(T)$ of degree $d$ let
$P^x:=\{F_1^x,\ldots,F_d^x\}\in\Part(E(M))$. 
Then $\nw(x)=\w(P^x)$. 

For an integer $k$ we let 
$\Pmtwk:=\{P\in\Part(E(M))\mid \w(P)\leq k\}$.
Let $\Ssing:=\big\{\{e\}\mid e \in E(M)\big\}\cup\{\emptyset\}$.

\begin{theo}\label{theo:mtw}
Let $M$ be a matroid with nonempty ground set $E(M)$
and let $k\geq 1$ be an integer. Then
\begin{enumerate}
	\item $(\Pmtwk,\Ssing)$ is a weakly submodular scenario on $E(M)$,
	\item $\mtw(M)\leq k$ if and only if the scenario 
		$(\Pmtwk,\Ssing)$ has a tree decomposition, and
	\item $(\Pmtwk,\Ssing)$ satisfies 
		Corollary~\ref{cor:biglist} (Characterising Tree Decomposable Scenarios).
	\item In particular, the scenario $(\Pmtwk,\Ssing)$ is monotone.
\end{enumerate}
\end{theo}

\proof
Statement 1 is straightforward to check using 
the fact that the rank function $\r$ is a connectivity function.
Statement 2~follows from Lemma~\ref{lem:move-to-leaves} and the definition
of $(\Pmtwk,\Ssing)$. 
Statement 3~follows from Corollary~\ref{cor:biglist}, together with 1 and 2.
\qed

\subsection{Tree-width of Graphs and Cycle Matroids}\label{subsec:twundaside}
In \cite{HlinenyW06,HlinenyW08}
it was shown that matroid tree-width of a cycle matroid $M[G]$
equals the tree-width of the corresponding graph $G$, provided the
graph has at least one edge.

We give a much shorter proof showing that
the matroid tree-width of a cycle matroid is not larger
than the tree-width of the corresponding
graph. After translating the definition of matroid tree-width
to graphs, the proof is purely graph theoretic.

We close the section by exhibiting the scenario
for the game characterising graph tree-width
(which is equivalent to the cops an robber game 
of \cite{seytho93}).

Let $G$ be a graph. 
A \emph{tree decomposition} of a graph $G=(V,E)$ is a pair 
$(T,B)$, consisting of a tree 
$T$ and a family $B=(B_t)_{t \in T}$ of subsets of $V$, the 
\emph{pieces} of $T$, satisfying:
	\begin{itemize}

	\item	For each $v \in V$ there exists $t \in T$, such that $v \in B_t$.
	(The node $t$ \emph{covers} $v$.)

	\item
	For each edge $e \in E$ there exists $t \in T$, such that $e \subseteq B_t$.
	(The node $t$ \emph{covers} $e$.)

	\item For each $v \in V$ the set
	$\{t \in T \mid v \in B_t \}$ is connected in $T$.
\end{itemize}
The \emph{width} of $(T,B)$ is defined as
$\w(T,B):= 
\max\big\{\left|B_t\right|-1\ \bigmid t\in T\big\}.$
The \emph{tree-width of $G$} is defined as
$
	\tw(G):= 
	\min\big\{\w(T,B)\ \bigmid (T,B) \text{ is a tree decomposition of }G\big\}.
$

By $M[G]$ we denote the corresponding
cycle matroid. For $F\subseteq E(G)$ we use $G\restrict F$ for 
denoting the subgraph $(V(G),F)$ of $G$. 
For $F\subseteq E(G)$ let $G-F:=G\restrict (E(G)\setminus F)$.  
Let $\c(G)$ denote the number of connected components of $G$. 
(Note that singletons play an important role when counting 
connected components.)
A \emph{vertex-free (VF) tree decomposition} \cite{HlinenyW06} 
of $G$ is a pair $(T,\tau)$, where $T$ is a tree, and
$
	\tau\colon E(G)\to V(T)
$
is a mapping. 
For a node $x\in V(T)$ of degree $d$, let again $T^x_1,\ldots,T^x_d$ 
denote the connected components of $T-x$, and let
$F^x_i:=\tau^{-1}(V(T^x_i))$.
The \emph{vertex free node-width} of $x$ is defined by
$
	\vfnw(x):=\left|V(G)\right|+(d-1)\cdot \c(G)-\sum_{i=1}^d\c(G-F^x_i).
$
The \emph{width} of the decomposition is the maximum vertex free node-width 
of the nodes of $T$, and the smallest width over all VF tree 
decompositions of $G$ is the \emph{VF tree-width} of $G$,
denoted by $\vftw(G)$. (The width of an empty tree is $0$.)
The following fact is not hard to prove (see \cite{HlinenyW06}).

\begin{fact}\label{fact:mtw-vf-tw} Let $G$ be a graph containing at least one edge.
Then $\mtw(M[G])=\vftw(G)$.
\end{fact}
The proof of the following theorem is an aside. It is much shorter than
the original proof in \cite{HlinenyW06}.

\begin{theo}\label{theo:tw-VFtw}
	Let $G$ be a graph with at least one edge. Then
	$\mtw(M[G])=\vftw(G)\le\tw(G)$. 
\end{theo}

\proof 
The equality follows from Fact \ref{fact:mtw-vf-tw}. Towards a proof of the inequality,
we may assume that $G$ is connected. 
Let $(T,B)$ be a \emph{small} tree decomposition of $G$ of width $k$,
i.e~there are no two nodes $s,t\in V(T)$, $s\neq t$, with $B_t\subseteq B_s$.
For every $e\in E(G)$ choose a node $t_e\in E(T)$ with
$e\subseteq B_{t_e}$. Define $\tau\colon E(G)\to V(T)$
by $\tau(e)=t_e$.
Note that since $(T,B)$ is small, the set $L(T)$ of leaves of $T$
is contained in the image of $E(G)$ under $\tau$, we have
$L(T)\subseteq \tau(E(G))$.

Let $x\in V(T)$, $d=\deg(x)$. We show that the node-width of $x$ is at most
$\left|B_x\right|-1$. Towards this, 
let $\partial F^x_i$ denote the set of
vertices incident with an edge in $F^x_i$ and an edge in 
$E(G)\setminus F^x_i$.
From $L(T)\subseteq \tau(E(G))$ it follows that $E(G-F^x_i)\neq\emptyset$ and hence
$1+\left|V(F^x_i)\setminus\partial F^x_i\right|\le \c(G-F^x_i)$, 
for $i=1,\ldots,d$. Moreover, (TD3)
implies that $\bigcup_{i=1}^d\partial F^x_i\subseteq B_x$.
The node-width of $x$ is
\begin{align*}
&\left|V(G)\right|+(d-1)\cdot \c(G)-\sum_{i=1}^d\c(G-F^x_i)
=\left|V(G)\right|+(d-1)\cdot 1-\sum_{i=1}^d\c(G-F^x_i)\\
&\le\left|V(G)\right|+d-1
-\sum_{i=1}^d\big(1+\left|V(F^x_i)\setminus\partial F^x_i\right|\big)
=\left|V(G)\right|-1
-\sum_{i=1}^d\left|V(F^x_i)\setminus\partial F^x_i\right|
=-1
\left|B_x\right|-1.
\end{align*}
The last equality follows from 
$
	V(G)= \bigcup_{i=1}^d(V(F^x_i)\setminus\partial F^x_i)
	\;\dot\cup\,B_x.
$
\qed

\medskip
As mentioned above, the inequality can actually be replaced by an equality.

\begin{fact}[Hlinen{\'y}, Whittle \cite{HlinenyW06,HlinenyW08}]\label{fact:tw-VFtw}
	Any graph $G$ with at least one edge satisfies
	$\mtw(M[G])=\vftw(G)=\tw(G)$. 
\end{fact}

Fact~\ref{fact:tw-VFtw} implies that Theorem~\ref{theo:mtw}
specialises to graphs (with at least one edge) 
in the expected way.
Nevertheless, let us make the scenario for tree-width of graph explicit.
Let $\Ssing:=\big\{\{e\}\mid e \in E(G)\big\}\cup\{\emptyset\}$.
Recall that for an integer $k\geq 1$, 
$\Ptwk=\big\{P\in\Part(E(G))\bigmid\left|\partial(P)\right| \leq k\big\}$.
Using Proposition \ref{prop:partial}, 
it is easy to check that the following holds.

\begin{theo}\label{theo:tw}
Let $G$ be a graph containing at leat one edge 
and let $k> 1$ be an integer. Then
\begin{enumerate}
	\item $(\Ptwk,\Ssing)$ is a weakly submodular scenario on $E(G)$,
	\item $\tw(G)\leq k$ if and only if the scenario 
		$(\Ptwk,\Ssing)$ has a tree decomposition, and
	\item $(\Ptwk,\Ssing)$ satisfies 
		Corollary~\ref{cor:biglist} (Characterising Tree Decomposable Scenarios).
\end{enumerate}
\end{theo}


\section{Conclusion}\label{sec:conclusion}
We introduced \emph{scenarios} and the \emph{captain and robber game} played on a
scenario. 
We proved that in all games on \emph{weakly submodular} scenarios,
the captain has a winning strategy, if and only if
the captain has a \emph{monotone} winning strategy,
i.e.~the games on weak submodularity scenarios are \emph{monotone}.
Extending ideas of \cite{ammanith08}, the proof uses 
search trees, tree decompositions of scenarios, 
and brambles in scenarios.

Our result implies monotonicity for a class of  
search games, that are equivalent to branch-width of submodular functions.
We obtain an exact characterisation for matroid tree-width
by a monotone game,
we obtain a monotone game equivalent to rank-width of graphs, and 
a monotone game characterising tree-width. 
Beyond this, our framework
also captures decompositions into `simple' subsets of the ground set.

We also included a proof showing that
the matroid tree-width of a graphic matroid is not larger
than the tree-width of the corresponding
graph. This proof is much shorter than the original proof~\cite{HlinenyW06}
and purely graph theoretic.

Moreover, with our framework it is easy to define a notion of \emph{branch-width}
for directed graphs, that is within a factor of $3$ of the
tree-width of a directed graph, as introduced in \cite{reed99}. 
We also obtain an exact game characterisation for tree-width 
of directed graphs.
These various applications give reason to believe that our framework
may also be useful in future research -- providing a tool for
producing game characterisations, that even come with a notion
of decomposition.

For scenarios that are not weakly submodular, it is still open
whether monotonicity holds. 
The games characterising hypertree-width \cite{gotleosca03} and
directed tree-width \cite{jrst01} are not monotone. 
Nevertheless, the monotone and the non-monotone variants
are strongly related \cite{jrst01,AdlerGG07}. In both cases, 
this relation is obtained via 
the notion of \emph{seperators}. Can we obtain similar
results for scenarios not satisfying
weak submodularity?
Can we extend the definition of scenarios and
obtain more insight into the open problems concerning monotonicity
of the games for DAG-width and Kelly-width \cite{KreutzerO08}?

The author thanks Tom\'a\v{s} Gaven\v{c}iak and Marc Thurley for comments on drafts
of this paper.
\bibliographystyle{plain}
\bibliography{games-archive}
\clearpage
\appendix 

\section{Appendix}
\label{appendix}

We restate and prove Theorem~\ref{theo:tdec-game}:
\begin{theo}\label{app:theo:tdec-game}
Let $(\mP,\mS)$ be a scenario on $A$.  

The pair $(\mP,\mS)$ has a tree decomposition if and only if
the captain has a winning strategy in the monotone captain and robber game
on $(\mP,\mS)$.
\end{theo}

We only sketch the proof. For graphs and hypergraphs, a detailed
proof of similar spirit can be found in \cite[Sect.~4]{Adler08}.

\medskip\noindent\emph{Proof sketch.} 
Let $(T,\tau)$ be a tree decomposition for $(\mP,\mS)$.
If $T$ has no internal nodes, then it is easy to see that the captain can win
by using the partition given by $\tau$. Otherwise,
the captain chooses an internal node $t$ and moves to $P_t$. Then the robber chooses a 
set $X\in P_t$. This set $X$ corresponds to the labels of the leaves of
exactly one component $T'$ of $T^{-t}$. The captain then chooses the 
neighbour $s$ of $t$ in $T'$ and moves to $P_s$. 
In this way the captain finally catches the robber in a leaf of $T$.

Conversely, suppose the captain has a winning strategy in the monotone
captain and robber game on $(\mP,\mS)$. Then the captain's strategy tree
gives rise to a tree decomposition of $(\mP,\mS)$. (The nodes of the strategy tree
correspond to the captain's partitions. The strategy tree
has the captain's first partition as a root, and a partition $\mP$
has a successor for every set $X\in\mP$ that the
robber can reach while the captain moves to $\mP$.) 
\qed

\medskip\noindent
We restate and prove Lemma~\ref{lem:make-exact}:

\begin{lem}\label{app:lem:make-exact}
	Let $A$ be a finite set, let $(\mP,\mS)$ be a weakly submodular scenario for $A$,
	and let $\mF\subseteq 2^A$. 
	
	If $(\mP,\mS)$ has a search tree compatible with $\mF$ 
	having at least one internal node, 
	then $(\mP,\mS)$ has an \textup{exact} search tree  
	compatible with $\mF$.
\end{lem} 

\proof Let $(T,l)$ be a search tree with at least one internal node 
for $(\mP,\mS)$, that is compatible with $\mF$.
We choose $l$ amongst all possible labelings 
such that the sum

\begin{equation}\label{eqn:maxsum}
\sum_{t\in V(T)\setminus L(T)}\;\sum_{X\in\pi_t}
\left|X\right|+\sum_{{\substack{s\in L(T)\\s'\in N(s)}}}\left|l(s,s')\right|
\end{equation}
is maximal. Suppose $st$ is a $2$-cycle of $T$ that is not exact.

If, say, $s$ is a leaf, then we can replace $l(s,t)$ by $l(t,s)^c$.
If neither of $s$ and $t$ is a leaf, then, by maximality of 
Sum (\ref{eqn:maxsum}), for every nonempty set 
$F\subseteq A\setminus(l(s,t)\cup l(t,s)^c)$ (such a set $F$ exists!) we have  
$(\pi_s)_{l(s,t)\to F}\notin \mP$. Hence, since $\mP$ is weakly submodular,
we can replace $\pi_t$ by $(\pi_t)_{l(t,s)\to F}$.
In both cases we obtain a search tree $(T,l')$ 
for ($\mP,\mS)$ compatible with $\mF$, 
where the size of Sum (\ref{eqn:maxsum}) 
is strictly increased, a contradiction.
\qed

\medskip\noindent
We restate and prove Theorem~\ref{theo:seachtree-tdec}:
\begin{theo}\label{app:theo:seachtree-tdec}
	Let $A$ be a finite set and let $(\mP,\mS)$ be a scenario for 
	$A$. 
	If the pair	$(\mP,\mS)$ has an exact search tree, then it has
	a tree decomposition.
\end{theo} 

\proof
We show that if the pair $(\mP,\mS)$ has an exact search tree $(T,l)$, 
then we obtain a tree decomposition $(T,\tau)$ for $(\mP,\mS)$ 
by letting $\tau(t):=l(s,t)$ for $t\in L(T')$. 
Since $(T,l)$ is a search tree for $(\mP,\mS)$, the mapping
$\tau$ is indeed a mapping from $L(T)$ to $\mS$.
By Fact~\ref{fact:exact-partition}, (TD1) is satisfied. 
If $T$ has at most one internal node, (TD2) is obviously satisfied as well.

For a $2$-arc $st$ let $T_t$ denote the subtree of
$T$ obtained by removing the arcs $(s,t)$ and $(t,s)$ from $T$, that 
contains $t$.
The following claim implies that Condition (TD2) holds.

\noindent
\emph{Claim.} All $2$-cycle $st$ with two internal nodes $s$ and $t$
satisfy 
\[l(s,t)=\bigcup_{{\substack{v\in L(T)\cap V(T_t)\\ 
u \in N(v)}}}l(u,v).\]
Towards proving the claim, let $T_t'$ be obtained from $T_t$ by adding 
vertex $s$ and the two arcs $(t,s)$ and $(s,t)$. Then $(T_t',l\restrict V(T_t'))$
is an exact search tree for $A$, and hence by Fact~\ref{fact:exact-partition},
the labels of the arcs entering the leaves of $T_t'$ form a partition
of $A$ and the claim follows.
\qed

\medskip\noindent
We restate and prove Theorem~\ref{theo:tdec-bdec}:

\begin{theo}\label{app:theo:tdec-bdec}
	Let $(\mP,\mS)$ be a scenario on $A$.
If $(\mP,\mS)$ has a tree decomposition, then
$(\mP,\mS)$ has a branch decomposition. 
\end{theo}

\proof
Let $(T,\tau)$ be a tree decomposition for $(\mP,\mS)$.
We turn $T$ into a cubic tree $T'$ by replacing 
every internal node $t$ of $T$, that has at least four neighbours 
$t_1,\ldots ,t_n$ ($n\geq 4$), by a cubic tree with leaves $t_1,\ldots, t_n$.
Identifying $L(T')$ with $L(T)$ in the obvious way, and 
using the fact that
$\mP$ is closed under coarser partitions, it is easy to see that
$(T',\tau)$ is a branch decomposition for $(\mP,\mS)$.
\qed

\medskip\noindent
We restate and prove Lemma~\ref{lem:move-to-leaves}:
\begin{lem}\label{app:lem:move-to-leaves}
Let $M$ be a matroid with $\mtw(M)\leq k$. There is
a tree decomposition $(T,\iota)$ for $M$ of width at most $k$
satisfying $\iota(E(M))\subseteq L(T)$.
\end{lem}

\proof 
Let $(T',\iota')$ be a tree decomposition for $M$ of width
at most $k$. For every element $e\in E(M)$
satisfying $\iota'(e)=t$, where $t\in V(T')$ is an internal node,
we create a new neighbour $t_e$ of $t$ and let
$\iota(e):=t_e$. It is easy to verify that in this way we obtain
the desired tree decomposition $(T,\iota)$ for $M$.
\qed

\end{document}